\documentclass[twocolumn,showpacs,preprintnumbers,amsmath,amssymb]{revtex4}
\usepackage{graphicx}
\input epsf
\usepackage{dcolumn}
\usepackage{bm}

\begin{document}

\title{The Destiny of Universes After the Big Trip}
\author{A.V. Yurov}%
\email{artyom_yurov@mail.ru}
 \affiliation{%
The Theoretical Physics Department, Russian State University of I.
Kant, A. Nevskogo str., 14, 236041,
 Russia.
\\
}
\date{\today}
\begin{abstract}
The big trip  can be describe with the help of the Wheeler-DeWitt
wave equation ${\hat H}\psi(w,a)=0$. The probability to find the
universe after big trip in the state with $w=w_0$ will be maximal
if $\partial\psi(w,a)/\partial w|_{w=w_0}=0$ for any values of the
scale factor $a$. It is  shown that this will be the case if and
only if $w_0=-1/3$. This fact allows one to suggest that  vast
majority of universes in multiverse must be in this state after
their big trips.



\end{abstract}

\maketitle

\section{\label{sec:level1}Introduction}

The cosmology nowadays is amazingly abundant with a new startling
solutions. Some of the most recent ones are the models with the
"Phantom fields" which result in the violation of the weak energy
condition (WEC) $\rho>0$, $\rho+p/c^2>0$ \cite{1}, \cite{2}, where
$\rho$ is the fluid density and $p$ is the pressure. Such phantom
fields, as follows from their quantum theory \cite{3}, should
inevitably be described by the scalar field with the negative
kinetic term. The through investigations shows that such fields
are apparently could not be considered as a fundamental objects.
However, it is possible that the Lagrangians with the negative
kinetic terms will appear as some kind of effective models, as it
happens in some models of supergravity \cite{4}, in the gravity
theories with highest derivatives \cite{5} and in field string
theory \cite{Aref'eva}, for example in model which is close to the
fermion NSR-string with regard for (GSO-)-sector (see also
\cite{Sen}). Finally, the "phantom energy" in the brane theory was
considered in \cite{7}, \cite{NPB}.
\newline
{\bf Remark 1}. Despite of all said above,
We can not be assured  that phantom energy is  only  effective model.
The  reason  is the existence of
''the crossing of the phantom divide line''. There are exact
solutions of Einstein equations which describes this crossing.
Moreover, this crossing is smooth and one can conclude that smooth
(de)-phantomization is the sufficiently general property of
Einstein equations. This effect was discover in \cite{tema1},
\cite{tema2} and it was interpreted in \cite{Andrianov}.
Therefore if we were (following \cite{Andrianov}) guided by a
belief that Einstein equations are more fundamental then the
concrete form of the Lagrangian for other fields then one can
conclude that ''the crossing of the phantom divide line'' is
possible and this is the new fundamental property of gravitation.

The particular interest to models with the phantom fields is
caused by their prediction of so-called "Cosmic Doomsday" alias
big rip \cite{1} (see also \cite{8}). In case of the phantom
energy we have $w=p/(c^2\rho)=-1-\epsilon$ with $\epsilon>0$.
Integration of the Einstein-Friedmann equation for the flat
universe results in
\begin{equation}
\begin{array}{cc}
\displaystyle{
a(t)=\frac{a_0}{\left(1-\xi t\right)^{2/3\epsilon}},}\\
\displaystyle{
\rho(t)=\rho_0\left(\frac{a(t)}{a_0}\right)^{3\epsilon}=\frac{\rho_0}{(1-\xi
t)^2}}, \label{1}
\end{array}
\end{equation}
where $\xi=\epsilon\sqrt{6\pi G\rho_0}$. We choose $t=0$ as the
present time, $a_0\sim 10^{28}$ cm and $\rho_0$ to be the present
values of the scale factor and the density. There, if
$t=t_*=1/\xi$, we automatically get the big rip.
\newline
{\bf Remark 2}. There are few ways to escape of future big rip singularity: (i) to
consider phantom energy just as some effective models (see above);
(ii) to use quantum effects to delay  the singularity
\cite{Nojiri}; (iii) to use new time variable such that the big
rip singularity will be point at infinity ($t\to\infty$)
\cite{Multiverse}; (iv) to avoid big rip via another cosmological ''Big'': big trip
(see below).

In \cite{9} Pedro F. Gonz\'alez-D\'iaz had shown that phantom
energy can results in achronal cosmic future where the wormholes
become infinite before the occurrence of the big rip singularity.
To show this lets consider the wormhole with the throat radius
$b_0=10^{-33}$ cm (Planck scale). It was shown in \cite{9} that if
$p =-(1+\epsilon)c^2\rho$ is a fluid's equation of state, then
\begin{equation}
c{\dot b}(t)=2\pi^2\epsilon GD\rho(t)b^2(t), \label{dotb}
\end{equation}
where $b(t)$ is the throat radius of a Morris-Thorne wormhole and
$D$ is dimensionless quantity. According to \cite{9} we can choose
$D\sim 4$ (see also \cite{10}). The equation (\ref{dotb})
describes the changing of the $b(t)$ with regard to the phantom
energy's accretion. Integration of the (\ref{dotb}) gets us
\begin{equation}
\frac{1}{b(t)}=\frac{1}{b_0}-\frac{2\pi^2\epsilon\rho_0 GD
t}{c(1-\xi t)}. \label{1b}
\end{equation}
Therefore at
\begin{equation}
{\tilde t}=\frac{c}{\epsilon(c\sqrt{6\pi G\rho_0}+2\pi^2\rho_0b_0G
D)} \label{tild}
\end{equation}
we get $b({\tilde t})=\infty$. As we can see ${\tilde t}<t_*$, and
therefore this universe indeed will be achronal before the
occurance of the big rip. In accord to \cite{9}, at $t>{\tilde
t}$, while in process of the phantom energy's accretion, the
wormhole becomes an Einstein-Rosen bridge which  can, in
principle, be used to escape from the big rip.
\newline
{\bf Remark 3}. In \cite{Faraoni} the capability of a phenomenon of big trip was
subjected a critic. In response article \cite{Est} the detailed answers to all objections of Faraoni  were given.

\section{Big trip and Wheeler-DeWitt equation}

The big trip is a cosmological process thought to occur in the
future by which the entire universe would be engulfed inside a
gigantic wormhole and might travel through it along space and time
(see Fig. 1).
\begin{figure}
\includegraphics[width=.9\columnwidth]{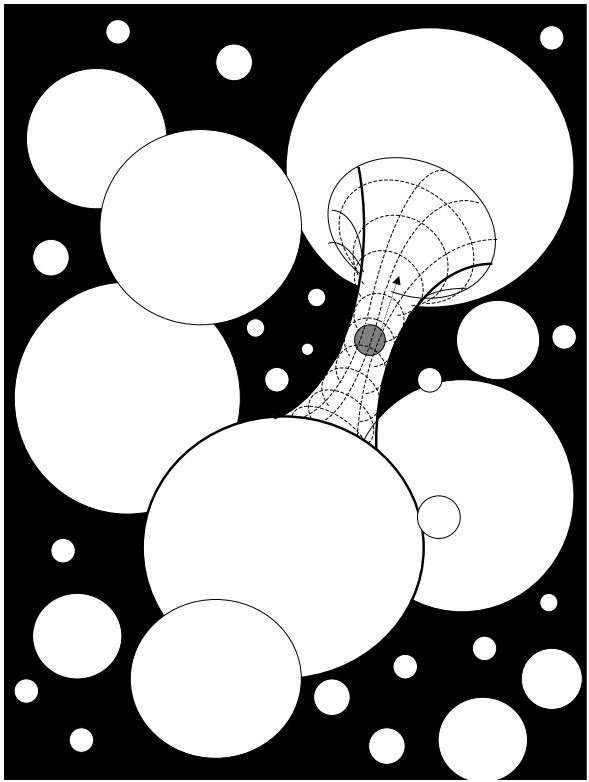}
\caption{\label{fig:epsart} Pictorial representation of the big
trip process when it is carried out by a single grown-up wormhole
within the framework of a multiverse picture. In this case the
universe does not travel along its own time but behaves like
though if its whole content were transferred from one different
larger universe to another, also larger universe.}
\end{figure}

In this article we'd like to present the possibility of new
cosmological ''Big'' - so called ''Big Meeting''.

Let us consider the spacetime manifold $M$ for a flat FRW universe
with metric:
\begin{equation}
ds^2=-N^2dt^2+a^2(t)d\Omega_3^2, \label{metr1}
\end{equation}
where $N$ is the lapse function and $a(t)$ is a scale factor.
Using the approach from the \cite{Pedro-Madrid} one suggest that
the parameter of equation of state $w=p/\rho$ ($c=1$) is
time-dependent one. We don't restrict ourselves by the condition
${\ddot w}=0$ but suppose that $\rho=\rho(w,a)$. If this is the
case then differentiating the Einstein-Friedmann equation
$H^2=8\pi G\rho/3$ with respect to $t$ one get new expression for
the scalar curvature ${\hat R}$:
\begin{equation}
{\hat R}=R+\frac{3}{{\dot a}a^2}\left(3(1+w){\dot
a}^3+a^3{\dot\rho}\right), \label{hatR2}
\end{equation}
where
$$
R=\frac{6({\dot a}^2+a{\ddot a})}{a^2},
$$
$$
{\dot\rho}=\frac{\partial\rho}{\partial a}{\dot
a}+\frac{\partial\rho}{\partial w}{\dot w}.
$$
We consider the universe filled with scalar field $\phi$ with
Lagrangian
$$
L=\frac{1}{2}\left(\partial_{\mu}\phi\right)^2-V(\phi)=p=w\rho.
$$
Therefore the action integral of the manifold $M$ with boundary
$\partial M$ has the form
\begin{equation}
S=\int_Md^4x\sqrt{-g}\left(\frac{{\hat R}}{16\pi
G}+w\rho\right)-\frac{1}{8\pi G}\int_{\partial M}
d^3x\sqrt{-h}{\rm Tr}{\hat K}, \label{action3}
\end{equation}
where ${\hat R}$ is the generalized Ricci curvature scalar
(\ref{hatR2}), $K$  is the conventional expression for the
extrinsic curvature, $g={\rm det}g_{\mu\nu}$, $h$ is the
determinant of the general threemetric on the given hypersurface
at the boundary $\partial M$. Since $\sqrt{-g}\sim Na^3$ then one
can integrate (\ref{action3}) over spatial variables and
substitute $t\to i\tau$ (with $8\pi G/3=1$) to reduce
(\ref{action3}) to Euclidean action:
\begin{equation}
I=\int d\tau
N\left[\frac{6aa'^3-3aF(a,a',w,w')}{a'N^2}+6a^3w\rho\right],
\label{I4}
\end{equation}
where
$$
F(a,a',w,w')=3(1+w)a'^3-a^3\left(\frac{\partial\rho}{\partial
a}a'+\frac{\partial\rho}{\partial w}w'\right),
$$
and ${}'=d/d\tau$. In the gauge where $N=1$ we have the
Hamiltonian constraint
\begin{equation}
H=\frac{\delta I}{\delta N}-(w+1)a^3\rho=0. \label{H5}
\end{equation}
The next step to the Wheeler-DeWitt equation is to define the
momenta conjugate to $a$ ($\pi_a$)  and $w$ ($\pi_w$) and redefine
the (classical) Hamiltonian (\ref{H5}) via $\pi_a$, $\pi_w$, $a$
and $w$. At last one must  introduce the following quantum
operators:
$$
{\hat \pi}_a=-i\frac{\partial}{\partial a},\qquad {\hat
\pi}_w=-i\frac{\partial}{\partial w},
$$
which allows one to obtain the  the Wheeler-DeWitt equation (WDE)
${\hat H}\psi=0$. Thus we have
\begin{equation}
4\left(a\frac{\partial\rho}{\partial
a}+2\rho\right)\frac{\partial^2\psi}{\partial
w^2}-4a\frac{\partial\rho}{\partial
w}\frac{\partial^2\psi}{\partial a\partial
w}=3\left(\frac{\partial\rho}{\partial w}\right)^2a^6(1+3w)\psi,
\label{WD6}
\end{equation}
with $\psi=\psi(a,w)$.

The WDE (\ref{WD6}) is differ from the WDE which was obtained in
\cite{Pedro-Madrid} because we didn't  use the condition ${\ddot
w}=0$ which allows one to simplify the actions (\ref{action3}),
(\ref{I4}) using  the rejection of corresponding surface terms. As
we shall see, the Eq. (\ref{WD6}) result in new ''Big'' in
cosmology - the ''Big Meeting''.

First at all, let consider the case $w=-1/3$. In this case the
right side of the (\ref{WD6}) will be zero. Moreover, if $w=-1/3$
then $\rho=\rho(a,w=-1/3)\sim a^{-2}$ therefore
$$
a\frac{\partial\rho}{\partial a}+2\rho=0,
$$
and the WDE (\ref{WD6}) is reduced to
\begin{equation}
\frac{\log a}{a}\frac{\partial^2\psi}{\partial a\partial w}=0.
\label{uh7}
\end{equation}
Using power series
$$
\psi(a,w)=\psi(a,-1/3)+\sum_{n=1}^{\infty}\frac{1}{n!}c_n(a)\left(w+\frac{1}{3}\right)^n,
$$
and (\ref{uh7}) we get $dc_1(a)/da=0$ thus
\begin{equation}
\frac{\partial \psi(a,w)}{\partial w}|_{w=-1/3}=c_1={\rm const}.
\label{8}
\end{equation}
On the other hand, the equation (\ref{uh7}) is invariant with
respect to transformation
\begin{equation}
\psi(a,w)\to\psi(a,w)-f_1(a)-f_2(w), \label{9}
\end{equation}
for arbitrary function $f_{1,2}$. Substituting (\ref{9}) into the
(\ref{8}) and choosing $df_1(w)/dw=c_1$ at $w=-1/3$ we get without
loss of generality
\begin{equation}
\frac{\partial \psi(a,w)}{\partial w}|_{w=-1/3}=0. \label{extr10}
\end{equation}
Therefore, in the case of general position, for any {\bf fixed}
$a$ the function $\psi(a,w)=\Phi_a(w)$ has the extremum at
$w=-1/3$. Since the function $\psi$ must be normalizable one, the
point $w=-1/3$ must be the point of maximum. In other words, the
probability distribution $|\psi(a,w)|^2$ for any given value of
the scale factor has the peak at $w=-1/3$.

One can prove this fact for the general position. Let consider the
equation (\ref{WD6}). We'd like to consider the solutions of this
equation such that
\begin{equation}
\frac{\partial \psi(a,w)}{\partial w}|_{w=w_0}=0, \label{extr11}
\end{equation}
for any given $a$. Since $w_0={\rm const}$ then
$\rho(a,w_0)=\rho_0=C^2a^{-3(w_0+1)}$ ($C={\rm const}$) and
\begin{equation}
a\frac{\partial\rho_0}{\partial a}=-3(w_0+1)\rho_0, \label{12}
\end{equation}
\begin{equation}
\frac{\partial\rho_0}{\partial w_0}=-3\log a\rho_0. \label{13}
\end{equation}
Besides
\begin{equation}
\frac{\partial^2\psi}{\partial a\partial
w}|_{w=w_0}=\frac{\partial}{\partial a}\left(\frac{\partial
\psi(a,w)}{\partial w}|_{w=w_0}\right)=0. \label{14}
\end{equation}
Substituting (\ref{12}), (\ref{13}) and (\ref{14}) into the
(\ref{WD6}) we get
\begin{equation}
(1+3w_0)\left(4a^{(-3(w_0+1)}\frac{\partial^2\psi}{\partial
w_0^2}+27C^2a^{-6w_0}\log^2a \psi\right)=0. \label{15}
\end{equation}
Using (\ref{15}) one can conclude that $w_0=-1/3$ or the following
equation must be hold
\begin{equation}
\frac{\partial^2\psi}{\partial w_0^2}=-\frac{27
C^2}{4}a^{-3(w_0-1)}\log^2a \psi. \label{16}
\end{equation}
The general solution of the (\ref{16}) has the form
\begin{equation}
\psi(w_0,a)=c_1J_0(z)+c_2Y_0(z), \label{solution17}
\end{equation}
where $c_{1,2}$ are arbitrary constants,
$z=\sqrt{3}Ca^{3(1-w_0)/2}$, $J_0$ and $Y_0$ are the Bessel
functions of the first and second kind. This function must be
normalizable one:
\begin{equation}
\int_0^{\infty} da|\psi(w_0,a)|^2<+\infty,
\label{integr}
\end{equation}
so one need to choose $c_2=0$ ($Y_0(z)\sim 2\log z/\pi$ at $z\to
0$ so if $c_2\ne 0$ then we get the divergence in the
(\ref{integr}) for $a\to 0$). Substituting (\ref{solution17}) into
the (\ref{14}) one get
$$
\frac{\partial\psi}{\partial w_0}=\frac{3\sqrt{3}
C}{2}a^{3(1-w_0)/2}\log a J_1\left(\sqrt{3} C
a^{3(1-w_0)/2}\right)=0,
$$
which will be the case for arbitrary $a$ if and only if $w_0=1$
and $C$ is the solution of the equation
$$
CJ_1(\sqrt{3}C)=0.
$$
But if $w_0=1$ then the wave function $\psi={\rm const}$ (and the
same will be the case for the density $\rho$ (see (\ref{WD6}))).
Such wave function will be non-normalizable one, thus, there is
only one way to comply with (\ref{extr11}) in framework of
normalizable wave function of the universe: to put $w_0=-1/3$.
{\em The end of the proof}.

One can ask about possibility of existence of another peaks
$M_*=(a_*,w_*)$ of solutions of the (\ref{WD6}), such that
$$
\frac{\partial\psi}{\partial
a}|_{M_*}=\frac{\partial\psi}{\partial w}|_{M_*}=0,
$$
and
\begin{equation}
\Delta=\left[\frac{\partial^2\psi}{\partial
w^2}\frac{\partial^2\psi}{\partial
a^2}-\left(\frac{\partial^2\psi}{\partial w\partial
a}\right)^2\right]_{M_*}>0. \label{det18}
\end{equation}
As we  shall see, it is possible to neglect these extremum. In
fact, one need to use the dimensionless variables in expressions
like $\log a$. Keeping in mind that our universe was born via big
trip from another, a paternal universe with the value of it's
scale factor $L$, one must to replace $\log a\to\log(a/L)$. On the
other hand, one can expect that the initial value of the scale
factor $a_i$ of the universe after the big trip will be $a_i\sim
L$. Let consider such universe with $w={\rm const}$. If we'd like
to estimate the probability to find this universe just after the
big trip (that is with $a_i\sim L\sim a_*$)  then one must use the
WDE (\ref{WD6}) which reduce to the simple form
$$
-4a^{-3(w+1)}(3w+1) \frac{\partial^2\psi}{\partial w^2}|_{M_*}=0.
$$
If $w\ne -1/3$ then (see (\ref{det18}))
$$
\Delta=-\left(\frac{\partial^2\psi}{\partial w\partial
a}\right)^2_{M_*}<0,
$$
and we have not extremum at all.

\section{Conclusion}

Therefore, as we seen, in the case of general position the
solution of (\ref{WD6}) and the probability distribution has the
peaks at $w=-1/3$ for any $a$ if we consider universes just after
big trip. This fact result in new possible ''Big'' in modern
cosmology which can be called as  ''Big Meeting''.  It means that
vast majority of universes in multiverse must be in the state with
$w=-1/3$ just after their big trips.

\acknowledgements

\noindent  I'd like to thank Pedro F. Gonz\'alez-D\'iaz for
important notices he made on the first draft of this paper.

$$
{}
$$
\bibliography{apssamp}

\vfill \eject

\end{document}